\documentstyle{l-aa}
\def\mabs{M$_{\rm abs}$}
\def\etal{et al.~}
\def\hii{H{\sc ii}}
\def\cbeta{$c_{\rm H\beta}$}
\def\micron{$\mu$m}
\def\kms{km s$^{-1}$}
\def\cmc{cm$^{-3}$}
\def\erg{erg s$^{-1}$ cm$^{-2}$ \AA$^{-1}$}
\def\ergs{erg s$^{-1}$}
\def\ergscm{erg s$^{-1}$ cm$^{-2}$}
\def\lsun{L$_{\odot}$}
\def\halpha{H$\alpha$}
\def\hbeta{H$\beta$}
\def\oi{[O\,{\sc i}]$\lambda$6300}
\def\oii{[O\,{\sc ii}]$\lambda$3727}
\def\oiiia{[O\,{\sc iii}]$\lambda$4959}
\def\oiiib{[O\,{\sc iii}]$\lambda$5007}
\def\nii{[N\,{\sc ii}]}
\def\niia{[N\,{\sc ii}]$\lambda$6548}
\def\niib{[N\,{\sc ii}]$\lambda$6583}
\def\sii{[S\,{\sc ii}]}
\def\siia{[S\,{\sc ii}]$\lambda$6716}
\def\siib{[S\,{\sc ii}]$\lambda$6731}
\def\Sii{[S\,{\sc ii}]$\lambda\lambda$6716,6731}
\def\oiiishb{[O\,{\sc iii}]/H$\beta$}
\def\niisha{[N\,{\sc ii}]/H$\alpha$}
\def\siisha{[S\,{\sc ii}]/H$\alpha$}
\def\infig#1#2#3{\epsfxsize=#3cm \centering{\mbox{\epsfbox{#2}}}}

\input epsf.sty

\begin{document}

   \thesaurus{02(11.19.3; 11.14.1; 11.01.2; 11.19.7; 09.08.1; 13.09.1)        
              }
   \title{Starbursts in barred spiral galaxies}

   \subtitle{III. Definition of a homogeneous sample of Starburst Nucleus 
Galaxies 
\thanks{Based on observations obtained at the 1.93 meter telescope of 
Observatoire de Haute-Provence operated by INSU (CNRS). Tables 1, 3, 4, 5 and 6 
are only available in electronic form at the CDS via anonymous ftp to {\tt 
cdsarc.u-strasbg.fr (130.79.128.5)} or via 
{\tt http://cdsweb.u-strasbg.fr/Abstract.html}}
             }

   \author{T. Contini \inst{1,2} \and
           S. Consid\`ere \inst{3} \and 
           E. Davoust \inst{1} 
           }

   \institute{
Observatoire Midi-Pyr\'en\'ees, UMR 5572, 14 Avenue E. Belin, F-31400 Toulouse, France
\and
School of Physics \& Astronomy, Tel-Aviv University, 69978 Tel-Aviv, Israel
\and 
Observatoire de Besan\c con, UPRES-A 6091, B.P. 1615, F-25010 Besan\c con Cedex, France
             }

   \offprints{T. Contini, contini@wise.tau.ac.il}
     
   \date{Received 30 September 1997; accepted 1 December 1997}

   \maketitle

   \begin{abstract}
This paper presents optical long-slit spectroscopic observations of 
105 barred Markarian IRAS galaxies. These observations are used to determine 
the spectral type (starburst or Seyfert) of emission-line regions in the 
nucleus and along the bar of the galaxies, in order to define a 
homogeneous sample of Starburst Nucleus Galaxies (SBNGs). 

Our selection criteria (ultraviolet excess, far infrared emission and barred 
morphology) have been very efficient for selecting star-forming galaxies, 
since our sample of 221 emission-line regions includes 82\% nuclear or 
extranuclear starbursts. The contamination by Seyferts is low (9\%). The 
remaining galaxies (9\%) are objects with ambiguous classification (\hii\ or 
LINER). 

The dust content and \halpha\ luminosity increase towards the nuclei of 
the galaxies. No significant variation of the electron density is found 
between nuclear and bar \hii\ regions. However, the mean \halpha\ luminosity 
and electron density in the bar are higher than in typical disk \hii\ 
regions. 
  
We investigate different mechanisms for explaining the excess of nitrogen 
emission observed in our starburst nuclei. There is no evidence for the 
presence of a weak hidden active galactic nucleus in our starburst galaxies. 
The cause of this excess is probably a selective enrichment of nitrogen in the 
nuclei of the galaxies, following a succession of short and intense bursts of 
star formation. 

Our sample of SBNGs, located at a mean redshift of $\sim$ 0.015, 
has moderate \halpha\ ($\sim 10^{41}$ \ergs) and far infrared ($\sim 10^{10}$ 
\lsun) luminosities. The types are distributed equally among early- and 
 late-type giant spirals with a slight preference for Sbc/Sc types because of 
their barred morphology. The majority (62\%) of SBNGs are isolated with no sign 
of gravitational interaction.  In terms of distance, luminosity and level of 
interaction, SBNGs are intermediate between \hii\ galaxies 
and luminous infrared galaxies. 
   
   
   \end{abstract}


\section{Introduction}

Our present knowledge of the properties of starburst galaxies, and of the 
physical conditions in nuclei and in extranuclear \hii\ regions of spiral 
galaxies, rests on the analyses of a rather limited number of data sets. The 
selection criteria for the various samples necessarily introduce biases which 
influence the derived properties and which might explain why these studies 
sometimes reach conflicting results, for example on the role of the bar or of 
gravitational interactions in triggering starbursts.  It is thus essential to 
multiply the observations of starburst galaxies, in order to make available 
the largest possible dataset for future analyses.   

The goal of the present paper is to present a new sample of starburst galaxies 
which differs markedly in its selection from other samples. We wanted to avoid 
selecting small metal-poor galaxies (such as blue compact dwarfs) 
preferentially found in objective prism surveys, monsters (such as mergers or 
ultraluminous infrared galaxies) selected because of their unusual shape or 
luminosity, and active galaxies (such as Seyferts), which are probably related 
to starbursts, but should be considered separately.  We thus devised selection 
criteria that would enable us to catch in our nets galaxies preferably
resembling rather ``normal" giant spiral galaxies.  We believe that we have 
been fairly successful, and this new sample has been subjected to an extensive 
multifrequency analysis (Contini 1996), the results of which have been or will 
be reported separately (see Sect.~\ref{END}). 

\section{The original sample}

\subsection{Selection criteria}

We defined three criteria for selecting a sample of galaxies likely to
contain the highest number of starburst galaxies.
The selected galaxies must possess:
\begin{enumerate}
\item an excess of blue and ultraviolet (UV) emission; this would indicate
the presence of a large number of young and hot stars. 
\item A strong far-infrared (FIR) emission; interstellar 
dust heated by the UV radiation field from the young and hot stars is
likely to give rise to such an emission.
\item A barred morphology; a bar is expected to enhance the gas flow 
toward the center of galaxies and may provide a mechanism for triggering 
nuclear starbursts.
\end{enumerate}

The Markarian survey contains 1500 objects selected for their excess of blue 
and UV emission among which about 1100 are galaxies. The catalog of Markarian 
galaxies is the largest sample of active galaxies known so far. The excess of 
blue luminosity and UV emission is the signature of either a high star 
formation rate or the presence of an Active Galactic Nucleus (AGN) in the 
central region of the galaxies. We extracted our sample from the extensive 
compilation of Markarian galaxies by \cite{MB86}~(1986) which lists valuable 
data, such as morphology, magnitude, radial velocity, radio (6, 11 and 21 cm 
continuum) and infrared (60 \micron) flux of the galaxies.  

The very successful IRAS mission, which covered about 98\% of the sky in 
four bands centered at 12, 25, 60 and 100 \micron\, has led to valuable 
information on the origin of the FIR radiation and its relation to the dust 
content of star forming regions. Interstellar dust is heated by absorption of 
a large fraction of UV radiation and this energy is reradiated 
in the FIR range. 
In most IRAS galaxies, the heating of interstellar dust is due to a high star 
formation rate and/or an AGN which provides a sufficiently strong ionizing 
flux to produce the observed FIR radiation. We thus selected those galaxies of 
the Markarian catalog which were detected by the IRAS satellite, and listed in 
two catalogs of IRAS sources, namely the Point Source Catalog (PSC) which 
contains all the bright point sources and the Faint Source Catalog (FSC) which 
contains fainter extended sources. 

With the first two criteria (UV- and FIR-bright), we expected to select a 
sample of galaxies with a high proportion of starburst galaxies. Our last 
criterion was to select galaxies with a barred morphology, because of their 
peculiar dynamics. The bar is thought to enhance the flow of molecular gas 
toward the center of spiral galaxies where a high concentration of this gas 
will trigger nuclear starbursts. To select galaxies with bars, we used the 
morphological classifications provided by \cite{MB86}~(1986) and LEDA
(Lyon-Meudon Extragalactic Database).

Finally, the combination of two activity criteria (Markarian + IRAS) and 
the presence of a bar defined a sample of 144 barred Markarian 
IRAS galaxies.   

This sample differs from others used for studying the starburst phenomenon in 
galaxies. Indeed, many works in this field are based on samples of small 
irregular or blue compact dwarf galaxies like ``\hii\ galaxies'' 
(e.g. \cite{TERLEVICHetal91}~1991), or samples of highly luminous infrared galaxies 
(e.g. \cite{VEILLEUXetal95}~1995), or even very heterogeneous samples of 
galaxies of mixed types.  
Our sample is of course not free from selection effects;
the first selection criterion excludes galaxies with little or no UV excess,
the second one discriminates against weaker starbursts, with less dust and
molecular gas, the third one favors late-type galaxies, and
the Markarian catalogue does not contain nearby star-forming galaxies
(such as M82 or NGC 253). 

\subsection{Global properties of the sample}

The global properties of the original sample are listed in Table~\ref{CAT}. 
The first column gives the number of the galaxy in the Markarian catalog. 
Columns 2 and 3 indicate the equatorial coordinates (equinox 1950). Cross 
identifications with other catalogs are given in column 4. 
The morphological properties of the galaxies, based on the classification 
system of \cite{DEVAUCOULEURSetal91}~(1991, RC3), are given in columns 5 to 
8. Column 6 indicates if the galaxy has ring structures and column 7 if 
it belongs to a multiple system. The inclination and heliocentric radial 
velocity are given in columns 9 and 10. The apparent blue magnitude, 
corrected for both Galactic and internal extinction with the method of 
\cite{DEVAUCOULEURSetal91}~(1991, RC3), and the absolute blue magnitude are 
reported in columns 11 and 12. All the above general properties come from 
LEDA. 

The IRAS data, reported in columns 13 to 20 of Table~\ref{CAT}, come
from the {\em Faint Source Survey} (\cite{BICAYetal95}~1995). These data are 
more accurate than those from the PSC or the FSC, and represent in fact the 
latest version of the IRAS catalog. The flux densities (in Jy) at 12, 25, 60 
and 100 \micron\ are reported in columns 14, 16, 18 and 20 
respectively. A quality code is assigned to each flux density in order to 
evaluate the reliability of the measurements. The code value is equal to 3 if 
the measurement is of high quality, to 2 if the quality is moderate and to 1 
if it corresponds to an upper limit (90\% confidence level). No measurements 
of the flux density at 100 \micron\ are reported with a quality code greater 
than or equal to 3, because of the strong contamination by interstellar cirrus 
at this wavelength.  Table~\ref{CAT} is given in electronic form only.
    
These data will be used in Sect.~\ref{SBNG} to derive the general properties 
of the SBNGs. 

\begin{table}
\caption[]{Global properties of the sample of barred Markarian galaxies}
\label{CAT}
\end{table}

\section{Observations}

The spectroscopic observations were obtained at the 1.93 meter telescope of 
Observatoire de Haute-Provence between April 1991 and January 1994. The data 
were acquired with the CARELEC spectrograph (\cite{LEMAITREetal90} 1990) and 
several types of CCD cameras (see Table~\ref{CCD}). The spectral resolution 
was 260 \AA/mm, which provides a spectral coverage of $\sim$ 3800 -- 7400 \AA\ 
with a resolution $\Delta\lambda \sim$ 15 \AA. The spectra were acquired 
under good photometric conditions with a typical seeing between 2 and 3\arcsec. 

During those nights, we also observed various spectrophotometric standard stars 
(see Table~\ref{CCD}) taken from the list given by \cite{MASSEYetal88}~(1988), 
except for BD~2626 (\cite{OG83}~1983), in order to flux calibrate the galaxy 
spectra. He comparison line spectra were obtained immediately before and after 
the galaxy integration in order to calibrate accurately the wavelength scale. 
The slit was usually aligned along the bar of the galaxy. For galaxies with 
multiple nuclei or for pairs of galaxies, we oriented the slit in order to 
cover the brightest knots. 

The observation log is given in Table~\ref{OBS}, which gives for each galaxy 
(column 1) the date of observation (column 2), the exposure time (column 3), 
the spectral range (column 4), the width and position angle of the slit 
(columns 5 and 6).  Table~\ref{OBS} is given in electronic form only.
During the allocated telescope time, we were able to 
observe 105 of the 144 galaxies of the sample. The remainder of the paper is 
based on the data collected for these 105 galaxies.

\begin{table}
\caption[]{Characteristics of CCD cameras and spectrophotometric standard stars used for each observing run}
{\scriptsize
\begin{flushleft}
\begin{tabular}{llcrl}
\hline
\hline
Observing run& \multicolumn{3}{c}{CCD camera}& standard \\
\cline{2-4}
 & Type & Size (pixels) & pixel size &   star  \\
\hline
April 1991 & RCA 3 & 323 $\times$ 512 & 30 \micron & GD140 \\
September 1991 & RCA 1 & 323 $\times$ 512 & 30 \micron & G191B2B \\
January 1992 & THX1 & 405 $\times$ 581 & 23 \micron &  \\
March 1992 & TK512 & 512 $\times$ 512 & 27 \micron & BD2626 \\
Feb.-March 1993 & TK512 & 512 $\times$ 512 & 27 \micron & GD140 \\
December 1993 & TK512 & 512 $\times$ 512 & 27 \micron & Hiltner 600\\
January 1994 & TK512 & 512 $\times$ 512 & 27 \micron & Hiltner 600 \\
\hline
\hline
\end{tabular}
\end{flushleft}
}
\label{CCD}
\end{table}

\begin{table}
\caption[]{Log of spectroscopic observations}
\label{OBS}
\end{table}

\section{Data reduction}
\label{REDU}

The spectroscopic data were reduced with the MIDAS package LONG. The
spectra were flux calibrated and corrected for foreground reddening
using the method described by \cite{CDC95}~(1995, hereafter paper I). 

In order to extract individual spectra of all (nuclear and extranuclear) 
emission-line regions for each galaxy spectrum, we plotted the spatial 
distribution of the \halpha\ emission line along the slit. We then identified 
distinct regions of emission and measured their dimension and position along 
the slit by fitting multi-Gaussian profiles to the \halpha\ spatial 
distribution. 

We extracted one-dimensional elementary spectra by adding the columns 
(spatial dimension) corresponding to the dimension of each emission-line 
region estimated above. One can note at this stage that 42 galaxies do not 
have any extra-nuclear emission-line region along the slit (i.e. along the 
bar). The elementary spectra were cleaned of radiation events (``cosmic rays") 
before further analysis. 

For measuring all the line parameters (fluxes, FWHM and equivalent widths), 
and for determining the internal reddening (\cbeta), we again followed the 
method described in paper I. For the latter determination, we adopted Case 
B of Balmer decrement ($I$(H$\alpha$)/$I$(H$\beta$) = 2.85) for classical 
\hii\ regions, and Case A  ($I$(H$\alpha$)/$I$(H$\beta$) = 3.10) for AGNs. 

We also estimated the continuum flux density, $f_{\lambda}$, of the
individual regions in three 
spectral bands labeled 
B ($\lambda_c$ = 4200 \AA\ and $\Delta \lambda$ = 300 \AA), 
V ($\lambda_c$ = 5400 \AA\ and $\Delta \lambda$ = 300 \AA) and 
R ($\lambda_c$ = 6800 \AA\ and $\Delta \lambda$ = 200 \AA) 
very close to Johnson's photometric system. 
In order to obtain the continuum values in a consistent and objective 
manner, we measured the spectra with an automated algorithm. We chose 
the wavelength range of the spectral bands in such a way as 
to exclude the regions containing emission lines. 

The measured emission lines are given in Table~\ref{MES}. 
The galaxy Markarian number is given in column 1. 
The label and code for the position of the 
emission-line regions (1 = nuclear, 0 = extranuclear) are given in columns 2 
and 3. The distance of the extranuclear regions from the nucleus, and the 
diameter of all regions, expressed in arcsec, are given in 
columns 4 and 5. The observed fluxes $F(\lambda)$ and 
equivalent widths of \hbeta, \halpha\ (both corrected for Balmer absorption), 
\oiiia, \oiiib, \niia, \niib, \siia\ and \siib\ are listed in columns 6 to 
21. 

The continuum measurements are listed in Table~\ref{CONT}. 
The galaxy Markarian number is given in column 1. The magnitudes
integrated along the slit ($m_{\rm B}$, 
$m_{\rm V}$ and $m_{\rm R}$) are given in columns 3, 4 and 5, where 
$m_{\lambda} = -2.5\log(f_{\lambda}) - 5\log(\lambda) - 2.41$, with 
$f_{\lambda}$ in units of \erg\ and $\lambda$ in \AA\ (\cite{MB93}~1993). We 
further used the continuum magnitudes to define two color indices: $B-V = 
m_{\rm B} - m_{\rm V}$ and $V-R = m_{\rm V} - m_{\rm R}$ which are reported in 
column 6 and 7. Tables~\ref{MES} and \ref{CONT} are given in electronic 
form only. 

\begin{table}
\caption[]{Measured spectrophotometric data. Observed intensities and 
equivalent widths of emission lines}
\label{MES}
\end{table}

\begin{table}
\caption[]{Measured spectrophotometric data. Magnitudes and color indices of 
the spectral continuum}
\label{CONT}
\end{table}

\section{Results}

The characteristics of the emission-line regions and the derived 
spectrophotometric parameters are given in Table~\ref{RES}. The galaxy (Mrk) 
number is given in column 1. The label and code for the position of the 
emission-line regions (1 = nuclear, 0 = extranuclear) are repeated
in columns 2 
and 3. The distance of the galaxy, estimated using a Hubble constant
of 75 kms$^{-1}$Mpc$^{-1}$, is given in column 4.  
The distance of the extranuclear regions from the nucleus, and the 
diameter of all regions, expressed in kpc, are given in 
columns 5 and 6. The reddening coefficient \cbeta\ is given in column 
7. The electronic density ($n_e$) 
is given in column 8.  The line-intensity ratios, \niib/\halpha, 
\Sii/\halpha\ and \oiiib/\hbeta, used for the spectral classification and 
corrected both for internal reddening and for Balmer absorption, are reported 
in columns 9, 10 and 11, followed by the results of the spectral 
classification in column 12. The \halpha\ flux [$I(\rm{H\alpha})$ in 
10$^{-14}$ \ergscm] is given in column~13 and the \halpha\ luminosity  
[$L(\rm{H\alpha}$) in \ergs] in column~14. In the subsequent analysis we 
only take into account the \halpha\ luminosities corrected for reddening, that 
is, those of galaxies for which we derived the reddening parameter \cbeta\ (see 
Sect.~\ref{REDDE}). For the others, we give lower limits on the \halpha\ flux 
and luminosity in Table~\ref{RES}, which is given in electronic form only. 

\begin{table}
\caption[]{Derived spectrophotometric parameters}
\label{RES}
\end{table}

\begin{figure*}
\vspace{-4cm}
\infig{12}{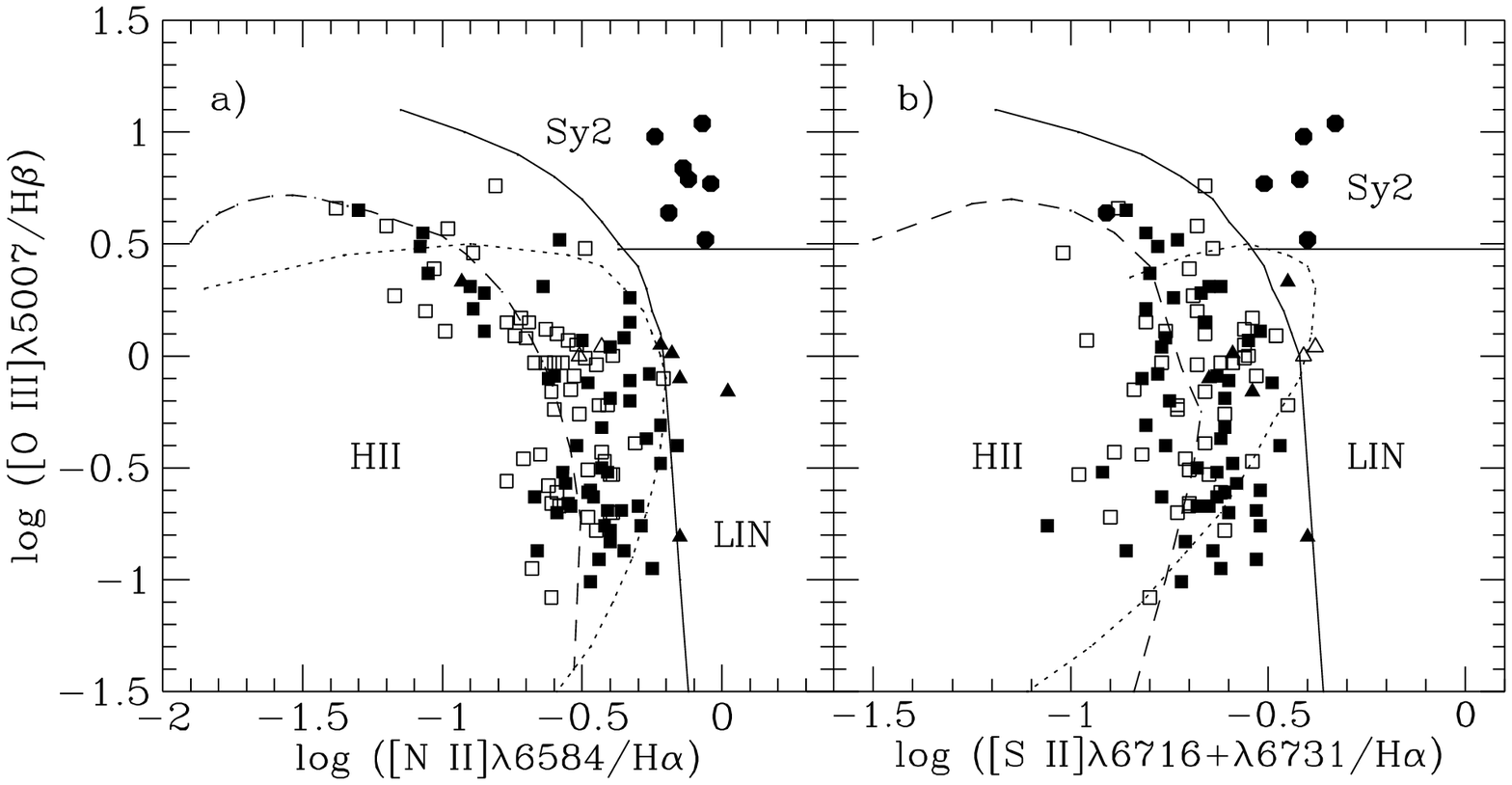}{16}
\vspace{-4cm}
\caption[]{Diagnostic diagrams, adapted from \cite{VO87}~(1987), showing 
{\it a}) log(\oiiib/\hbeta) vs. log(\niib/\halpha) and 
{\it b}) log(\oiiib/\hbeta) vs. log(\Sii/\halpha). The symbols are : 
{\it filled squares} = starburst nuclei, 
{\it open squares} = extranuclear \hii\ regions, {\it filled circles} = 
Seyfert 2 nuclei and {\it triangle} = objects with an ambiguous 
classification (\hii\ or LINER, see text). 
In each diagram two curves ({\it solid lines}) 
separate the emission-line regions in three groups: starburst galaxies or 
\hii\ regions (lower left), Seyfert 2 galaxies (upper right) and LINERs 
(lower right). The {\it dashed curve} denotes the theoretical 
models of disk \hii\ regions of \cite{MRS85}~(1985). 
Also shown is the photoionization model of \cite{SK95}~(1995) 
({\it dotted curve})}
\label{TSPEC}
\end{figure*}

\subsection{Spectral Classification}

We used~diagnostic diagrams (Baldwin, Phillips \& Terlevich 1981, \cite{VO87} 1987) for 
identifying the various ionization mechanisms responsible for the 
emission-line regions of our galaxies.  In these diagrams, one can distinguish 
regions photoionized by hot and young stars (i.e. \hii\ regions) from those 
photoionized by a harder radiation field, such as that of an AGN or a LINER, 
using several ratios between low-ionization and Balmer emission lines. 

At this stage, it is necessary to clarify the definition of the different types 
of starburst galaxies which will be mentioned in this paper. One can first 
separate \hii\ galaxies and starburst galaxies by the properties of the 
host galaxy. \hii\ galaxies are mainly metal-poor dwarf irregular or 
blue compact galaxies with low dust content 
(e.g. \cite{F80}~1980, \cite{K83}~1983 and \cite{TERLEVICHetal91}~1991). 
They contain many giant \hii\ regions with high-excitation spectra, with 
properties very close to those of extragalactic giant \hii\ regions 
distributed in the disk of nearby spiral galaxies (e.g. \cite{MRS85}~1985). 
\hii\ nucleus galaxies (\cite{S82}~1982, \cite{K83}~1983, 
\cite{KKB89}~1989, \cite{HFS97a}~1997a) and Starburst Nucleus Galaxies 
(SBNGs; \cite{B83}~1983, \cite{COZIOLetal94}~1994) are defined as 
galaxies with \hii\ regions in their nuclei. They 
are more massive and chemically evolved spiral galaxies, with a large 
population of old and evolved stars and a huge quantity of dust 
(\cite{C96}~1996). The low-excitation spectra of SBNGs reflect the higher 
metallicity observed in spiral galaxies, and especially in their nucleus, 
when compared to irregular or compact \hii\ galaxies. A last distinction is 
made between \hii\ nuclei and SBNGs on the basis of their nuclear \halpha\ 
luminosity, \hii\ nuclei being fainter (L(H$\alpha$) $\la 10^{40}$ \ergs) 
than SBNGs. In terms of star formation, \hii\ nuclei represent the 
low-luminosity end of SBNGs (\cite{C96}~1996).

The results of our spectrophotometric analysis are displayed in 
Fig.~\ref{TSPEC}\ which shows the location of the emission-line regions in 
two diagnostic diagrams of \cite{VO87}~(1987). In each diagram, the continuous 
curve, empirically derived by \cite{VO87}~(1987), separates starburst nuclei 
and \hii\ regions where the gas is assumed to be ionized by young stars, from 
AGNs where the main ionizing source is thought to be an accretion disk around 
a black hole (e.g. \cite{R84}~1984) which produces a power law spectrum.  We 
make a further distinction among AGNs between objects of high (\oiiishb\ $>$ 
3) and low (\oiiishb\ $\leq$ 3) excitation (horizontal line, 
\cite{SO81}~1981). The first group represents the classical Seyfert 2 galaxies 
while the LINERs fall in the second group. We did not use the original 
criteria for defining a LINER (\cite{H80}~1980) because measurements of \oi\ 
and [O\,{\sc ii}]$\lambda$3727 were often not available. 
For the same reason, we 
did not make use of the \oi/\halpha\ ratio to distinguish between the 
different sources of ionization. 

\begin{table}
\caption[]{Spectral classification}
\begin{flushleft}
\begin{tabular}{llrrr}
\hline
\hline
\multicolumn{2}{c}{Spectral Type$^{\rm a}$} & \multicolumn{2}{c}{Emission line region} & {\bf Total} \\
\cline{3-4}
                &                 & Nuclear   & Extranuclear                 &  \\
\hline
\multicolumn{2}{l}{\hii\ \dotfill} &  70  (67\%) &  81  (69\%) & 151  (68\%) \\
      & SBNG                    &  65  (62\%) &   ...          &    ...         \\
      & \hii G                  &   5   (5\%) &   ...          &    ...         \\[1mm]
\multicolumn{2}{l}{AGN \dotfill} &  20  (19\%) &    ...         &  20   (9\%) \\
      & Sey 1                   &  12  (11\%) &  ...           &    ...         \\
      & Sey 2                   &   8   (8\%) &   ...          &     ...        \\[1mm]
\multicolumn{2}{l}{\hii, LINER?} &  11   (10\%) &   8   (7\%) &  19   (9\%) \\[1mm]
\multicolumn{2}{l}{Uncertain}   &   4   (4\%) &  27  (24\%) &  31  (14\%) \\[1mm]
\hline
\multicolumn{2}{l}{\bf Total \dotfill}       & 105 (100\%) & 116 (100\%) & 221 (100\%) \\
\hline
\hline
\noalign{\smallskip}
\noalign{$^{\rm a}$ SBNG = Starburst Nucleus Galaxy; \hii G = \hii\ galaxy; 
\hii, LINER? = ambiguous classification between \hii\ and LINER (see text for 
details); Uncertain = classification based on only one emission line ratio 
(\niisha\ or \siisha)}
\end{tabular}
\end{flushleft}
\label{SCLASSI}
\end{table}

One should note that the classification process is not always unambiguous, for 
at least two reasons. First, the two conditions involving the low-ionization 
lines ([N\,{\sc ii}] and [S\,{\sc ii}]) do not always hold simultaneously. 
This reflects the empirical nature of the diagnostic diagrams as well as the 
possibility that one line ratio is enhanced or depressed with respect to the 
other one as a result of, for instance, selective abundance variations (see 
Sect.~\ref{NII}). Second, large measurement uncertainties may be associated 
with any given line intensity ratio. Thus, one should evaluate each object 
individually, taking all of these factors into consideration, before a 
classification can be assigned to it. When more than one classification is 
consistent with the data, both are given, with the more likely one listed 
first (column 12 of Table~\ref{RES}). An ambiguous spectral classification 
(between \hii\ and LINER) arises for 11 nuclear (Mrk 90, 271, 332, 353, 593, 
617, 874, 1180, 1200, 1291 and 1485) and 8 extranuclear (Mrk 712-3, 814-1, 
814-4, 1086-3, 1302-1, 1363-1, 1363-3 and 1433-3) regions.  

The result of the spectral classification is summarized 
in Table~\ref{SCLASSI}. We found that 70 nuclear regions (67\% of the sample) 
have spectra characteristic of photoionization by hot stars, 
i.e. are classified as starburst nuclei (62\%) or \hii\ galaxy (5\%). 
Four galaxies (Mrk 21, 271, 446 and 1452) were classified as SBNGs
using only one emission-line ratio ([N\,{\sc ii}]/\halpha\ or 
[S\,{\sc ii}]/\halpha), their classification is thus rather uncertain. 
AGN emission lines were observed in 20 nuclei (19\%) including 12 Seyfert~1 
galaxies (11\%) and 8 Seyfert~2 galaxies (8\%). 
Among the 116 extranuclear regions, 81 (69\%) are \hii\ regions. Here, the 
classification is uncertain for 27 regions (24\%) and ambiguous between
\hii\ and LINER for 8 regions (7\%) for the same reason as explained above. 

\begin{figure}
\vspace{-0.5cm}
\infig{12}{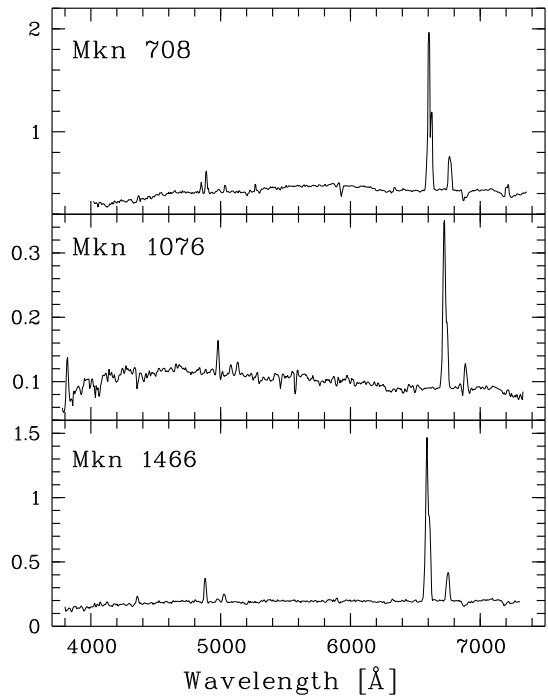}{8.8}
\vspace{-3.5cm}
\caption[]{Optical spectra of starburst nuclei. Intensities are in 
$10^{-14}$ \erg}
\label{stb2}
\end{figure}

\begin{figure}
\vspace{-0.5cm}
\infig{12}{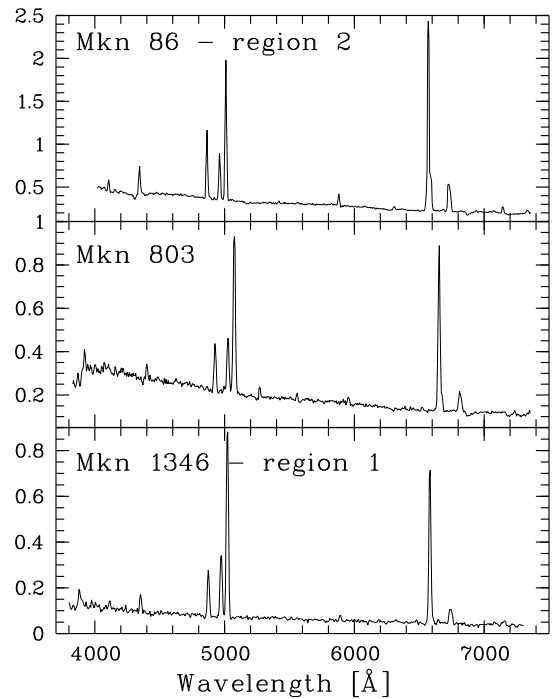}{8.8}
\vspace{-3.5cm}
\caption[]{Optical spectra of \hii\ galaxies showing strong emission 
lines and a high excitation (\oiiishb\ $>$ 3) compared to starburst nuclei. 
Intensities are in $10^{-14}$ \erg}
\label{H2}
\end{figure}

\begin{figure}
\vspace{-0.5cm}
\infig{12}{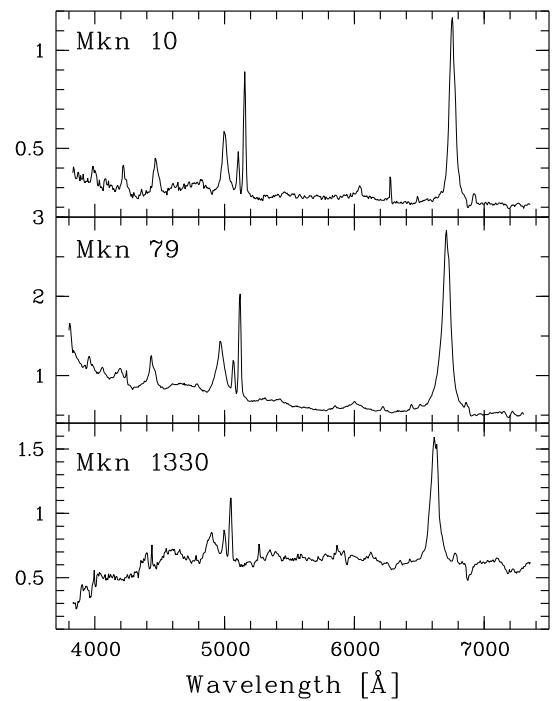}{8.8}
\vspace{-3.5cm}
\caption[]{Optical spectra of Seyfert 1 nuclei. The broad Balmer emission 
lines (\halpha, \hbeta, etc), compared to the narrower forbidden lines (e.g. 
[O\,{\sc iii}]), allow a clear and rapid classification of these AGNs. 
Intensities are in $10^{-14}$ \erg}
\label{SEY1}
\end{figure}

\begin{figure}
\vspace{-0.5cm}
\infig{12}{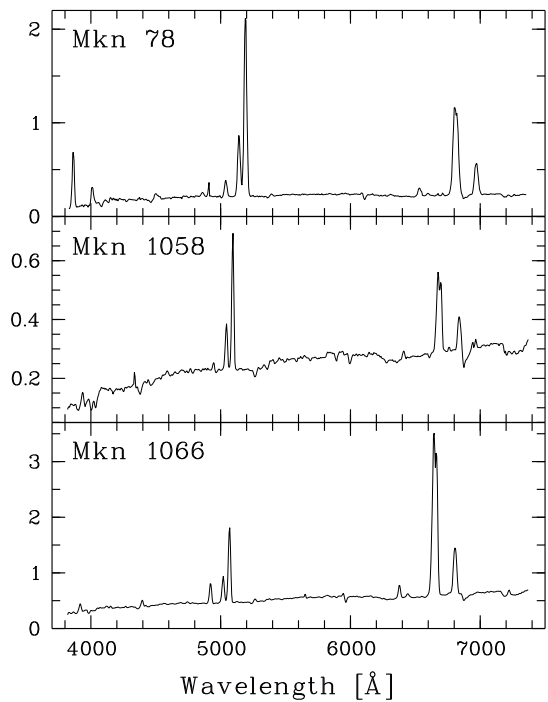}{8.8}
\vspace{-3.5cm}
\caption[]{Optical spectra of Seyfert 2 nuclei. In these AGNs, the Balmer 
lines have the same width as the forbidden lines and the excitation is high 
(\oiiishb\ $>$ 3). Intensities are in $10^{-14}$ \erg}
\label{SY2}
\end{figure}

\begin{figure}
\vspace{-0.5cm}
\infig{12}{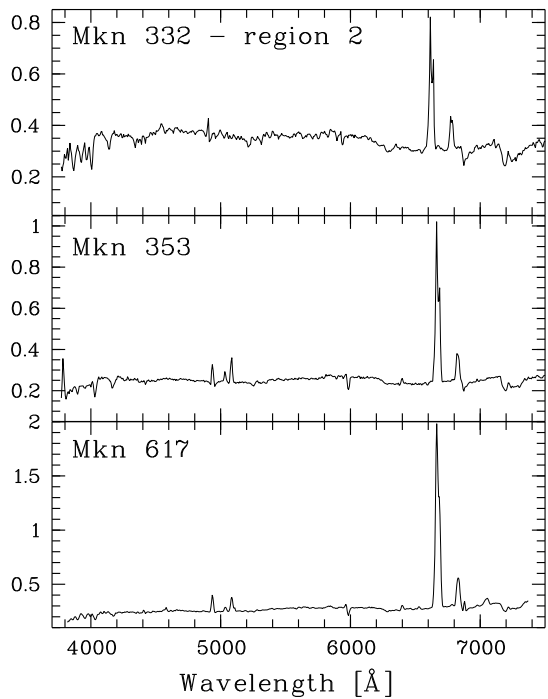}{8.8}
\vspace{-3.5cm}
\caption[]{Examples of objects with an ambiguous classification 
between \hii\ and LINER. Intensities are in $10^{-14}$ \erg}
\label{TRAN}
\end{figure}

Representative spectra of each spectral class are shown in Fig.~\ref{stb2}\ 
(starburst nuclei), Fig.~\ref{H2}\ (\hii\ galaxies), Fig.~\ref{SEY1}\ 
(Seyfert~1 nuclei), Fig.~\ref{SY2}\ (Seyfert~2 nuclei), and Fig.~\ref{TRAN}\ 
(objects with an ambiguous classification). 

\subsection {Colors}

We corrected for reddening the spectral continuum colors $(B - V)$ and 
$(V - R)$ of the individual regions
using the reddening coefficient \cbeta\ and assuming that the 
interstellar extinction applies in the same way for the stellar population 
and the ionized gas in emission-lines regions.

\begin{figure}
\vspace{-0.5cm}
\infig{12}{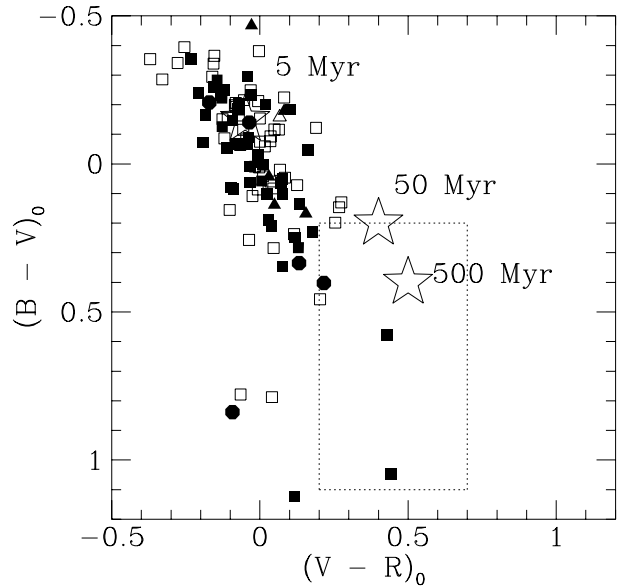}{8.8}
\vspace{-1cm}
\caption[]{Dereddened color-color diagram. The symbols 
have the same meaning as in Fig.~\ref{TSPEC}. The dotted rectangle indicates 
the position of a sample of 500 normal galaxies (\cite{BW95}~1995). The 
predictions of stellar population synthesis models of \cite{LH95}~(1995) are 
indicated by the big stars for 5, 50 and 500 Myr}
\label{COLOR}
\end{figure}

A color-color diagram with the dereddened color indices $(B - V)_0$ and 
$(V - R)_0$ is shown in Fig.~\ref{COLOR}. The extranuclear \hii\ regions 
and starburst nuclei are well mixed in this diagram, indicating identical 
stellar populations born during the same star formation episode.  
We first compare the colors of these starburst regions 
to the total color indices of  ``normal'' galaxies. 
It comes as no surprise that they are
much bluer than quiescent galaxies whose position is indicated 
by the dotted rectangle in Fig.~\ref{COLOR}; the vast majority of our 
starburst regions
are located outside this rectangle traced by about 500 
normal galaxies (\cite{BW95}~1995). We then use the predictions of stellar 
population synthesis models of \cite{LH95}~(1995) to estimate the age of the 
stellar population which dominates the spectral continuum of these
starburst regions.
It appears clearly that the colors observed in the star-forming regions 
are well fitted by a very young stellar population with an age lower than 
50 Myr.  

\subsection {Reddening}
\label{REDDE}

\begin{figure*}
\infig{12}{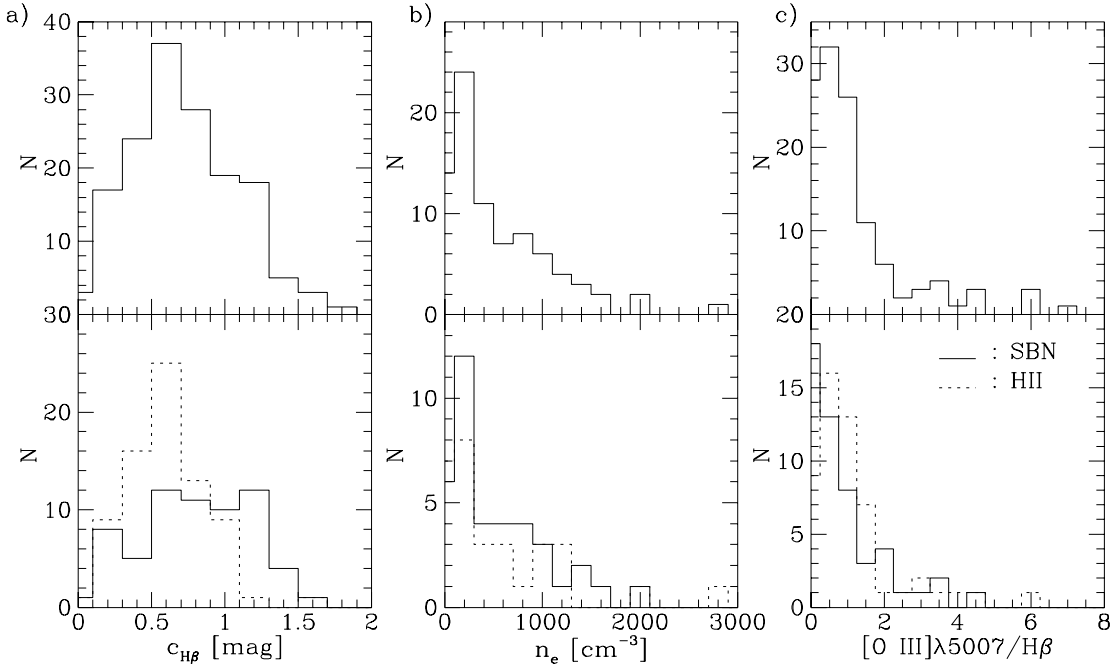}{16}
\vspace{-7cm}
\caption[]{Distribution of 
a) the extinction coefficient \cbeta, 
b) the electron density $n_{\rm e}$ and 
c) the excitation parameter \oiiib/\hbeta\ 
for all the emission-lines regions ({\it top}), 
for the starburst nuclei (SBN) and the extranuclear \hii\ 
regions (HII) ({\it bottom})
}
\label{HISTSPEC}
\end{figure*}

The distribution of reddening (\cbeta) in the different emission-line
regions is given in column 7 of Table~\ref{RES} and shown in
Fig.~\ref{HISTSPEC}a. The amount of reddening is larger
in starburst nuclei (\cbeta\ = 0.81$\pm$0.38)
than in extranuclear
\hii\ regions (0.61$\pm$0.28). When compared to other samples of
starburst
galaxies, the mean extinction coefficient derived in our sample is
slightly
larger than in nearby \hii\ nuclei (\cbeta\ $\sim$ 0.42,
\cite{HFS97a}~1997a)
and disk \hii\ regions (\cbeta\ $\sim$ 0.29, \cite{KKB89}~1989) but low
compared to a sample of luminous infrared starburst galaxies (\cbeta\
$\sim$
0.99, \cite{VEILLEUXetal95}~1995).

The amount of reddening has not been estimated for 13 nuclear and 35
extranuclear emission-line regions because of the weakness or absence
of \hbeta\ emission in their spectra. In all these objects
we detect a relatively strong \halpha\ emission; they are thus probably
highly obscured. Note that in a few objects (nucleus of Mrk~52, extranuclear 
regions of Mrk~489 and 712), the observed Balmer decrement is significantly 
less than the theoretical value; we assigned an internal extinction of 
zero to these objects.

\subsection {Electron density}

We derived the electron density ($n_e$) from the reddening-corrected 
\siia/\siib\ line ratio using the analytical relation given by 
\cite{O89}~(1989). A 20\% uncertainty in the \siia/\siib\ flux ratio 
corresponds to an uncertainty of about 100 cm$^{-3}$ in the determination of 
$n_e$.  

Figure~\ref{HISTSPEC}b shows the distribution of the electron densities for 
the different emission-line regions.  The mean value of the electron density 
is nearly the same for nuclear starbursts (560$\pm$240 \cmc) and extranuclear 
\hii\ regions (770$\pm$500 \cmc). These mean values are higher than those 
derived for nearby \hii\ nuclei (180$\pm$200 \cmc, \cite{HFS97a}~1997a), disk 
\hii\ regions ($\sim$ 140 \cmc, \cite{KKB89}~1989) and luminous infrared 
starburst galaxies ($\sim$ 280 \cmc, \cite{VEILLEUXetal95}~1995).  
    
\subsection {Line-intensity ratios}

\subsubsection {\oiiishb\ as an excitation parameter}

Two types of starburst galaxies can be distinguished based on their 
level of excitation: SBNGs show low-excitation spectra (\oiiishb\ $<$ 3, 
see Fig.~\ref{stb2}) whereas \hii\ galaxies show high-excitation spectra 
(\oiiishb\ $\geq$ 3, see Fig.~\ref{H2}).

A quick inspection of Fig.~\ref{TSPEC}\ indicates that the upper left 
region of the diagnostic diagrams contains only a few data points, reflecting 
the fact that essentially all the emission-line regions classified as starbursts
have a relatively low excitation level (\oiiishb\ 
$\leq$ 3). This is to be contrasted with Figs. 1--3 of \cite{VO87}~(1987) 
where this region of the diagrams is populated with 
extranuclear \hii\ regions and the low-metallicity \hii\ galaxies 
from the sample of \cite{F80}~(1980). 

The distribution of the excitation parameter is shown in 
Fig.~\ref{HISTSPEC}c. 
Our sample is obviously deficient in \hii\ galaxies, since only 
five starburst galaxies (Mrk 86, 412, 803, 860 and 1346) have an 
excitation parameter \oiiishb\ $\geq$ 3. 
The properties of these galaxies (listed in Table~\ref{CAT}) indicate that 
they are mainly small,
irregular and low-mass galaxies with a low dust content, confirming the 
general trend of this class of starburst galaxies (\cite{C96}~1996). Our sample 
thus contains a vast majority of starburst nuclei located in 
more massive and chemically evolved galaxies than \hii\ galaxies because 
of their higher frequency of past bursts of star formation
(\cite{C96}~1996). Note however that SBNGs are still 
in the process of formation because of their lower metal content compared 
to ``normal'' spiral galaxies (\cite{COZIOLetal97a}~1997a). 
Figure~\ref{HISTSPEC}c also shows that 
the mean excitation parameter is slightly higher in extranuclear \hii\ 
regions (\oiiishb\ $\sim$ 0.72) than in starburst nuclei 
(\oiiishb\ $\sim$ 0.50), reflecting the negative abundance gradient from 
the nucleus to the outer parts of spiral galaxies.  

\begin{figure}
\vspace{-0.5cm}
\infig{12}{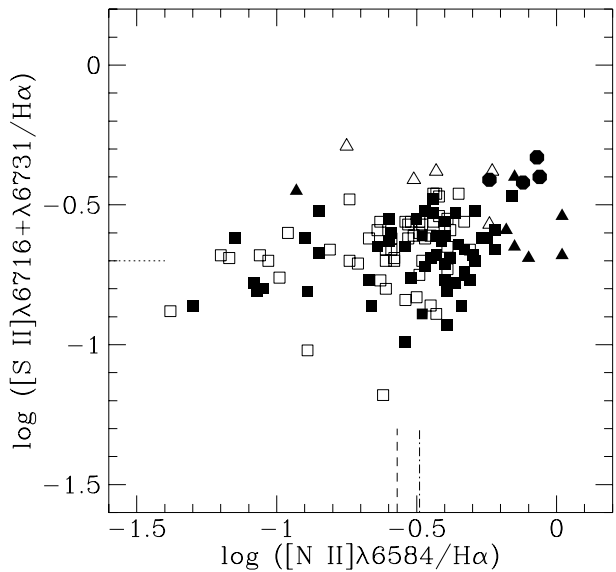}{8.8}
\vspace{-1cm}
\caption[]{Diagram of \siisha\ as a function of \niisha. The symbols 
have the same meaning as in Fig.~\ref{TSPEC}. The mean value of \niisha\ is 
higher in starburst nuclei ({\it dotted-dashed line}) than in extranuclear 
\hii\ region ({\it dashed line}) whereas the mean values of \siisha\ are
identical ({\it dotted line})} 
\label{N2S2}
\end{figure}

Three Wolf-Rayet galaxies are present in our sample. The optical 
spectrum of this subset of starburst galaxies shows broad emission lines 
from Wolf-Rayet stars around $\sim$ 4700 \AA, the brightest line being 
He\,{\sc ii}$\lambda$4686. While Mrk 52 and Mrk 710 were 
already known Wolf-Rayet galaxies and included in the catalog of 
\cite{C91}~(1991), a new one, Mrk 712, was discovered in the
sample (paper I).

\subsubsection {Excess of \niisha\ in starburst nuclei}
\label{NII}

Excess emission of [N\,{\sc ii}] has been reported in 
samples of emission line 
galaxies, such as SBNGs (\cite{COZIOLetal97b} 1997b) or \hii\ 
nuclei (\cite{HFS97a} 1997a) with a mean ratio of 
$\log$(\niisha) 0.2 to 0.3 dex higher than the mean ratio observed 
in disk \hii\ regions or predicted by normal \hii\ regions models 
(e.g. \cite{MRS85}~1985). 

In the diagnostic diagram of Fig.~\ref{TSPEC}a, one can see that, 
for a given 
excitation parameter, 
our nuclear starbursts tend to have stronger \niib\ emission 
than extranuclear \hii\ regions, but the difference is rather small
($<$ 0.1 dex) compared to other samples of starburst nuclei cited above.

This excess of nitrogen emission was first noted by \cite{S82}~(1982) in 
the nuclei of ``normal'' galaxies and confirmed later in a sample of 
``\hii\ region-like'' nuclei by \cite{KKB89}~(1989) who proposed 
the presence of a hidden weak AGN to account for this excess of 
low-ionisation emission line. AGNs indeed produce a harder ionizing 
radiation field than young O- or B-type stars. These high-energy photons 
create an extensive partially ionized zone from which low-ionization 
emission lines, such as \nii, \sii\ and [O\,{\sc i}] originate. 

To be sure that no hidden AGN is located in the starburst nuclei of 
our sample, we compare in Fig.~\ref{N2S2} two ratios of 
low-excitation emission lines, that of \siisha\ and that of 
\niisha. In the presence of a harder ionizing spectrum, both 
ratios should increase and a correlation would appear. One can  
clearly see that there is no such relation between the two ratios,  
neither for the nuclear starbursts nor for the
extranuclear ones, which are well mixed in this 
diagram. In fact, the mean value of \siisha\ is nearly identical for 
starburst nuclei ($\sim$ 0.20) and extranuclear \hii\ regions 
($\sim$ 0.21). The presence of a weak 
AGN in the nuclei of our starburst galaxies is also excluded because of 
the weakness of \oi\ in their spectrum: only $\sim$ 25\% 
of our nuclear spectra show this emission line with intensities similar 
to those observed in normal \hii\ regions (\cite{VO87}~1987).

Alternative explanations, like collisional excitation by shocks 
(\cite{KKB89} 1989) or very hot O type stars in metal-rich environments 
(\cite{FT92} 1992, \cite{S92} 1992), have been suggested as ionization sources 
to account for the excess of nitrogen emission in galactic nuclei. However, 
both suggestions fail to reproduce our observations, because they also imply
an increase of other low-ionization emission lines like [S\,{\sc ii}] and 
[O\,{\sc i}]. 

We have investigated whether the impact of dust on the thermal properties 
of \hii\ regions would provide a better explanation.
The dust content in our sample is not 
negligible since all our galaxies are IRAS sources (one of our selection 
criteria).
Calculations by \cite{SK95}~(1995) indicate that the 
influence of dust on the emergent optical spectrum of \hii\ regions can be 
quite appreciable in high-metallicity ($Z > Z_{\odot}$) environments, as is 
the case in many galactic nuclei. In Fig.~\ref{TSPEC}, we compare our data 
to the results of the photoionization model of \cite{SK95}~(1995) which 
incorporates the effects of dust and is calculated for a stellar effective 
temperature of 45 000 K.  The predicted line strengths do not 
provide a good match for all the observations in our starburst nuclei.
The model accounts reasonably well for the \niisha\ ratios observed in high 
metallicity 
(\oiiishb\ $\leq$ 0.5) nuclei and for regions very close to the transition 
limit between \hii\ and LINERs, but this appears to be accidental,
since the predicted \siisha\ ratio does not match our observations.
These theoretical results might simply be the consequence of the selective 
initial element abundances, since \cite{SK95}~(1995) arbitrarily assumed 
an enhancement of nitrogen abundance, with a secondary component scaling 
as $Z^2$, while other elements are in solar proportions. The results of 
the photoionization model of \cite{SK95}~(1995) might thus follow from 
this selective abondance introduced {\it ad hoc} in the model.

\begin{figure*}
\infig{12}{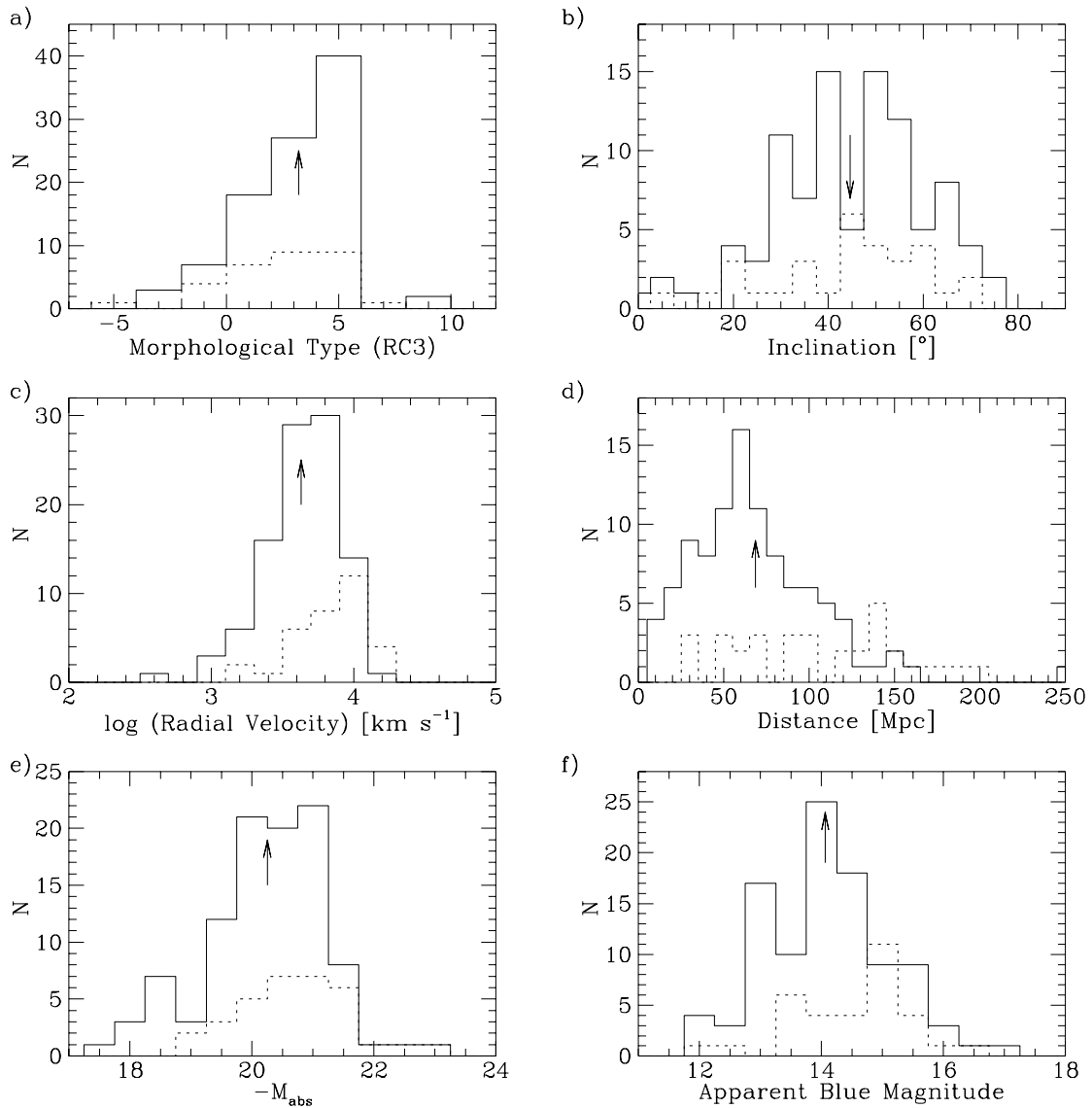}{16}
\vspace{-0.5cm}
\caption[]{Distribution of 
a) morphological type from RC3, 
b) inclination, 
c) heliocentric radial velocity, 
d) distance,
e) absolute blue magnitude and 
f) apparent blue magnitude for 
SBNGs ({\it solid line}) and AGNs ({\it dotted line}). Mean values are 
indicated by vertical arrows}
\label{HISTCAT}
\end{figure*}

An enhancement of nitrogen in starburst nuclei is certainly the most 
reliable explanation to account for the moderate excess of nitrogen emission 
in our sample of galaxies. 
Such selective chemical enrichment of nitrogen has been observed in the 
interstellar medium of some nearby starburst galaxies, like NGC~5253 
(\cite{WR89}~1989, \cite{KOBULNICKYetal97}1997) where N-enriched regions are 
found in the vicinity of young starbursts with a large population of massive 
Wolf-Rayet stars (\cite{SCHAERERetal97}1997). Moreover, chemical evolution 
models of galaxies (e.g. \cite{MMT94}~1994) predict an enhancement of nitrogen 
abundance after a succession of short and intense bursts of star formation, 
which has certainly been the case in starburst nuclei (\cite{C96}~1996). 

\section{Global properties of SBNGs}
\label{SBNG}

\subsection{Morphological types, distances, magnitudes and environment}

The global properties of our sample of SBNGs 
are displayed in Fig.~\ref{HISTCAT}, which presents the distribution 
of morphological types (according to RC3), inclinations, heliocentric 
radial velocities, distances, blue apparent and absolute magnitudes of 
the galaxies, using the values tabulated in Tables~\ref{CAT} and \ref{MES}. 
In these diagrams, we show the global properties of SBNGs after removing 
the five \hii\ galaxies and galaxies with ambiguous nuclear spectral 
classification. The global properties of AGNs (Seyfert 1 and 2, LINERs) are 
also shown for comparison.

The mean inclination of the galaxies is $\sim$ 45~\degr$\pm$16, both for 
SBNGs and AGNs. This value is comparable to that derived for barred 
galaxies with \hii\ nuclei (\cite{HFS97b}~1997b). The heliocentric radial 
velocity of SBNGs is predominantly ($\sim$ 93\%) lower than or equal to 10,000 
\kms\ with mean values around 4300 and 7000 \kms\ for SBNGs and AGNs 
respectively. The SBNGs are located at a mean distance of 68$\pm$38 Mpc, 
slightly farther than nearby \hii\ nuclei ($\sim$ 22 Mpc; 
\cite{HFS97b}~1997b). The mean distance derived for the AGNs is 107$\pm$50 
Mpc. The mean value of the apparent blue magnitude is 14, with a large 
proportion ($\sim$ 90\%) of giant spiral galaxies (\mabs\ $\leq -19.5$). The 
mean value of the absolute magnitude ($\sim -20.2$) is identical for our 
sample of SBNGs and for nearby \hii\ nuclei (\cite{HFS97b}~1997b). 

The SBNGs are equally distributed among early-type (S0-a to Sb; 55\%) and 
late-type (Sbc to Sm; 42\%) galaxies with only 3\% elliptical galaxies 
as is expected in a sample of starburst galaxies. In fact, the proportion 
of SBNGs increases with the morphological type and reaches a maximum of 40\% 
for Sbc/Sc galaxies. The distribution of AGNs is on the contrary more 
uniform from S0 to Sc. One can also note the deficiency of SBNGs
with morphological types later than Sc, which confirms the low 
contamination of our sample by low-luminosity blue compact and irregular 
\hii\ galaxies.

Coziol, Barth \& Demers (1995) found a majority of early-type galaxies 
among their sample 
of SBNGs, but this result should be considered with caution because only 39\% 
of their galaxies are morphologically classified. On the contrary, 
\cite{HFS97b}~(1997b) found a majority of late-type galaxies (62\% of Sc-Sm) 
among their sample of \hii\ nuclei. They note however that this effect is 
pronounced in barred galaxies (65\%) whereas unbarred galaxies with \hii\ 
nuclei are equally divided between early and late types. This may explain the 
relatively high frequency of Sc galaxies found in our sample of SBNGs since 
all our galaxies are barred. 

\begin{figure}
\infig{12}{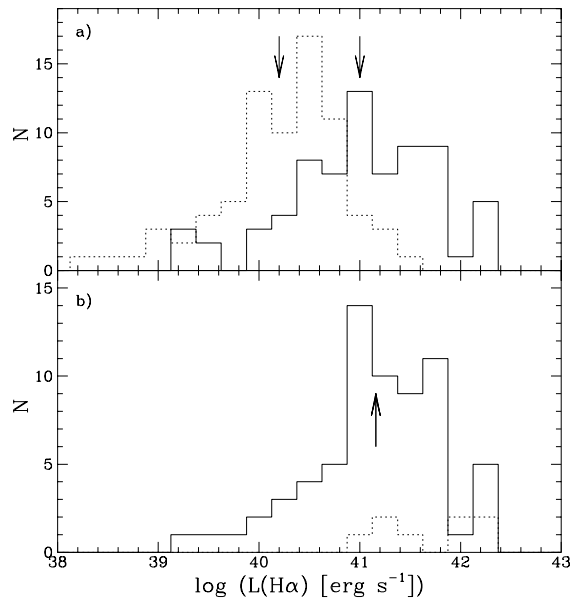}{8.8}
\vspace{-0.5cm}
\caption[]{
a) Distribution of reddening-corrected \halpha\ luminosity 
for individual nuclear ({\it solid line}) and extranuclear \hii\ region 
({\it dotted line}).
b) Distribution of total (nuclear and extranuclear) reddening-corrected 
\halpha\ luminosity for the SBNGs ({\it solid line}) and AGNs ({\it dotted 
line}). Mean values are indicated by vertical arrows 
}
\label{HISTLUM_HAL}
\end{figure}

One of the most popular ideas for explaining powerful starbursts is that they 
must occur preferentially within galaxies undergoing gravitational 
interactions. We thus performed an analysis of environment and level of 
interaction for all our galaxies using CCD images (\cite{CONTINI96}~1996). We 
found that the majority (62\%) of SBNGs are isolated galaxies. Only two 
galaxies (Mrk 617 and 960) are advanced mergers, 6\% of the galaxies belong to 
close pairs (projected distance $\leq$ 1\arcmin) and 23\% to wide pairs 
(projected distance $\leq$ 15\arcmin\ and $\Delta V \leq$ 300 \kms). More than 
half of the SBNGs do not show any sign of past or present gravitational 
interaction. Asymmetries in the bar or spiral arms are observed in only 32\% 
of the galaxies; among them, 12\% have multiple bright knots along the bar. 
This does not indicate that bars are necessary for triggering starbursts
in the absence of interactions; other samples of
(barred {\it and} unbarred) SBNGs (\cite{COZIOLetal97b}~1997b) and
\hii\ galaxies (\cite{TT95}~1995) have also been found to
contain a low proportion ($\sim$ 20 to 25\%) of interacting galaxies.
Interactions are more frequent among luminous infrared galaxies. The
level of interactions increases with the FIR luminosity, the proportion of
mergers reaching a maximum among ultra-luminous infrared galaxies
(\cite{VEILLEUXetal95}~1995).

\subsection{\halpha\ luminosity}

\begin{figure*}
\infig{12}{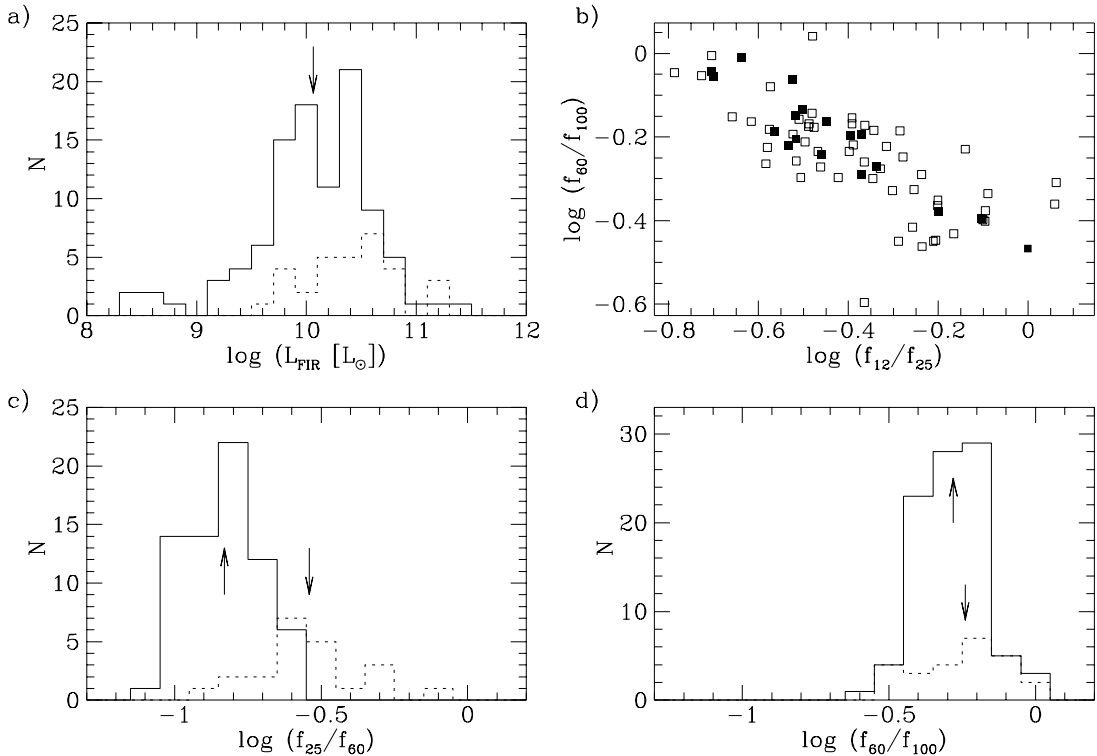}{16}
\vspace{-5.5cm}
\caption[]{
a) Distribution of total FIR luminosity for the SBNGs ({\it solid line}) 
and AGNs ({\it dotted line}). 
b) IRAS color-color diagram for SBNGs ({\it open squares}) and AGNs 
({\it filled squares}), 
c) and d) Distributions of $f_{25}/f_{60}$ and $f_{60}/f_{100}$ IRAS colors for 
SBNGs ({\it solid line}) and AGNs ({\it dotted line}). Mean values are 
indicated by vertical arrows 
}
\label{HISTLUM_FIR}
\end{figure*}

In this section and the next, we compare the distribution of the \halpha\ and 
FIR luminosities of our sample of SBNGs with those derived for other samples 
of starburst galaxies. A more detailed and quantitative discussion on the 
relation between \halpha, blue and FIR luminosities, 
and on the distribution of 
\halpha\ equivalent widths in terms of star formation history and age of the 
starbursts is given in Contini, Consid\`ere \& Davoust (in preparation).   

It appears clearly in Fig.~\ref{HISTLUM_HAL}a that the average \halpha\ luminosity 
[log(L(H$\alpha$)/\ergs)] is higher (by a factor $\sim$ 10) in starburst 
nuclei (41.0$\pm$0.7) than in extranuclear \hii\ regions (40.2$\pm$0.6). 
However, the \halpha\ luminosity of the \hii\ regions, which are 
mainly located along the bar of our 
galaxies, is higher than that of typical disk \hii\ regions 
(39.5; \cite{KKB89}~1989). The \halpha\ luminosities estimated in our 
starburst nuclei are typical of starburst galaxies (40.7; \cite{B83}~1983). 
They are clearly higher than in nearby \hii\ nuclei (39.2; 
\cite{HFS97a}~1997a) but slightly lower than in starbursts in luminous 
infrared galaxies (42.0; \cite{VEILLEUXetal95}~1995).  Contrary to what occurs 
in nearby \hii\ nuclei (\cite{HFS97a}~1997a), we do not find any significant 
difference between the total \halpha\ luminosities of early-type 
(41.3$\pm$0.4) and late-type (41.0$\pm$0.7) SBNGs. 

In terms of \halpha\ luminosity, our sample of SBNGs is thus intermediate 
between nearby \hii\ nuclei and luminous infrared galaxies. These luminosities 
are typical of starburst galaxies and comparable to other samples of 
SBNGs (i.e. \cite{COZIOLetal94}~1994).

As shown in Fig.~\ref{HISTLUM_HAL}b, the total (nuclear and extranuclear) \halpha\ luminosity derived for SBNGs (41.2$\pm$0.6) is slightly lower than 
that derived for AGNs (41.7$\pm$ 0.5). These luminosities are very close to those observed in other samples of Seyfert galaxies ($\sim$ 42.0; 
\cite{DR88}~1988, \cite{VEILLEUXetal95}~1995). 

\subsection{FIR properties}

We computed the FIR luminosities of the galaxies from the IRAS flux densities 
at 60 and 100 \micron\ (Table~\ref{CAT}) using the following relation which 
approximates well the total FIR luminosity between 42 and 122 \micron\ 
(\cite{HELOUetal88}~1988) 

\begin{equation}
\log(L_{\rm FIR}) = 5.5378 + 2\log(D) + \log(2.58f_{60} + f_{100})
\end{equation}
where the flux densities at 60 and 100 \micron\ are expressed in Janskys, 
$D$ is the distance of the galaxy in Mpc and $L_{\rm FIR}$ is the FIR 
luminosity in solar units.

The distribution of FIR luminosities [log(L$_{\rm FIR}$/\lsun)] is shown in 
Fig.~\ref{HISTLUM_FIR}a. Our sample of SBNGs has moderate FIR luminosities 
(10.1$\pm$0.5), slightly higher than those observed in samples of \hii\ nuclei 
($\sim$ 9.4) and \hii\ galaxies ($\sim$ 8.9), similar to other samples of 
SBNGs ($\sim$ 9.9) (see \cite{C96}~1996 and references therein) but rather low 
compared to samples of luminous ($\sim$ 11) or ultra-luminous ($\sim$ 12) 
infrared galaxies (\cite{VEILLEUXetal95}~1995). There is no significant 
difference between the average FIR luminosity of the SBNGs and AGNs 
(10.4$\pm$0.4). 
The AGNs in our sample follow the trend observed in infrared-bright galaxies, 
their proportion increases with FIR luminosity and reaches a maximum of 62\% for 
$\log(L_{\rm FIR}/L_{\odot}) \geq 12$ (\cite{VEILLEUXetal95}~1995). 
  
We did not use the fluxes at 12 and 25 \micron\ to compute the FIR 
luminosities, because of the strong contribution of non-thermal radiation 
to dust heating at these wavelengths. This well-known phenomenon 
(e.g. \cite{MNS85}~1985, \cite{DEGRIJPetal85}~1985) is illustrated in 
Figs.~\ref{HISTLUM_FIR}c,d where we show the distribution of the two IRAS 
colors, $f_{25}/f_{60}$ and $f_{60}/f_{100}$, for both SBNGs and AGNs. 
While no difference is seen in the distribution of $f_{60}/f_{100}$ 
(mean value $\sim$ 0.55) for AGNs and starburst galaxies, a clear excess 
of emission at 25 \micron\ is observed for AGNs [$\log(f_{25}/f_{60}) = 
-0.5\pm 0.2$] when compared to SBNGs [$\log(f_{25}/f_{60}) = -0.8\pm 0.1$]. 
This indicates that the infrared emission at short wavelengths (12 and 25 
\micron) is mainly due to a ``warm dust" component heated by the non-thermal 
ionizing radiation from AGNs. Figure~\ref{HISTLUM_FIR}b also shows that, regardless 
of the spectral classification, there is a clear tendency for galaxies 
with ``warmer" 60/100 colors to have ``cooler" 12/25 colors, illustrating 
the need for a multicomponent model to describe the nature of IRAS 
infrared emission (e.g. \cite{H86}~1986). Such a model requires the presence 
of a ``warm dust" component of infrared emission associated with star 
formation regions, and a ``cool dust'' component associated with the neutral 
interstellar medium.

\section {Summary and conclusion}
\label{END}

We presented optical long-slit spectroscopic observations of 105 
barred Markarian IRAS galaxies. These observations were 
mainly used to assign a spectral type (nuclear starburst, \hii\ region, Seyfert~1 
or~2) to each emission-line region along the slit, and to define a homogeneous 
sample of starburst nuclei and extranuclear \hii\ regions. 

Our selection criteria (UV excess, FIR emission and barred morphology) have 
been very efficient for selecting star-forming galaxies, since our sample of 
221 emission-line regions includes 82\% nuclear or extranuclear 
starbursts. The contamination by AGNs is low (9\%), with 13 Seyfert~1 
and 8 Seyfert~2 nuclei. The remainder are objects with ambiguous classification 
between \hii\ and LINER. 

Our sample of star-forming galaxies contains only 5 \hii\ galaxies, 
characterized by a high excitation parameter, equivalent to a low 
metallicity. In fact, our sample is strongly biased towards energetic 
starbursts located in the nuclear or extranuclear regions of more massive
and chemically evolved galaxies than \hii\ galaxies. 
Three Wolf-Rayet galaxies (Mrk 52, 710 and 712) are also included in our 
sample. 

We first compared the physical properties of the starburst nuclei to those of 
extranuclear \hii\ regions distributed along the bar and to those of typical 
disk \hii\ regions (\cite{KKB89}~1989). The amount of reddening, and hence the 
dust content, increases towards the nucleus of the galaxy. The same trend is 
also observed for the \halpha\ luminosity; the highest star formation rates 
are observed in the nuclei of the 
galaxies. We also found that the mean \halpha\ 
luminosity of the bar \hii\ regions is higher than that of typical disk \hii\ 
regions, probably because we are dealing with starburst galaxies.
We did not observe any significant variation of the electron density 
in the nuclei and bar \hii\ regions, but the measured values are higher (by a 
factor of about 3) than in typical disk \hii\ regions.  The excitation 
parameter, \oiiishb, generally decreases from the center outwards, reflecting 
the negative metallicity gradient observed in barred 
(Consid\`ere, Contini \& Davoust, in preparation) and 
ordinary spiral galaxies. 
  
We investigated different mechanisms for explaining the excess of nitrogen 
emission observed in starburst nuclei, which however is low compared to that 
estimated in nearby \hii\ nuclei and other SBNGs. There is no evidence for the 
presence of a weak hidden AGN in the nuclei of our starburst galaxies, as 
suggested by \cite{KKB89}~(1989) for explaining the excess of nitrogen in other 
samples of such galaxies. The most likely explanation is a selective 
enrichment of nitrogen in the nuclei of galaxies, following a succession of 
short and intense bursts of star formation. 

The properties of our sample of SBNGs are very much like those of other 
samples of SBNGs (\cite{COZIOLetal97b}1997b) and starburst galaxies 
(\cite{B83}~1983) located at a redshift of $\sim$ 0.01 -- 0.02 and with nearly 
the same \halpha\ and FIR luminosities. The host galaxies are distributed 
equally among early- and late-type giant spirals with a slight preference for 
Sbc/Sc types in our sample, a selection effect caused by the presence of a 
bar. The majority of SBNGs are isolated with no sign of gravitational 
interaction, contrary to the opinion that starbursts in massive galaxies are 
produced by gravitational interactions. This result suggests that, in the 
majority of spiral galaxies, bursts of star formation may depend on internal 
mechanisms, rather than on gravitational interactions.  

SBNGs are intermediate between low-mass irregular \hii\ galaxies 
(\cite{TERLEVICHetal91}~1991) or nearby \hii\ nuclei (\cite{HFS97a}~1997a) 
and luminous or ultra-luminous infrared galaxies (\cite{VEILLEUXetal95}~1995). 
The former are closer and intrinsicaly less luminous in \halpha\ and in the 
FIR whereas the latter are farther and more luminous, both in \halpha\ and 
FIR, with a high proportion of interacting galaxies. 

This spectrophotometric dataset on starbursts has been used to determine the 
age and star formation rate of the starbursts (Contini, Davoust \& 
Consid\`ere, in preparation) and, 
together with millimetric observations of the molecular gas, to establish 
the presence of molecular gas outflows in the nuclei of barred starburst 
galaxies (\cite{CONTINIetal97a}1997a). New spectrophotometric observations 
of a subset of this sample, with better signal-to-noise ratio and including 
the \oii\ emission line, have been used to derive the metallicity of the 
nebular gas and its gradient along the bar of the galaxies 
(Consid\`ere, Contini \& Davoust, in preparation), and have revealed yet another Wolf-Rayet galaxy.

The detailed spectroscopic analysis of some galaxies of the sample (Mrk 710, 
712 and 799), combined with CCD imaging and  observations of molecular clouds 
and atomic hydrogen has given rise to new results concerning the population of 
massive stars and the starburst properties in Wolf-Rayet galaxies 
(\cite{CONTINIetal97b}1997b). The analysis of the optical 
and CO velocity fields of Mrk 799 is in progress. 

The link between the morphological and dynamical parameters of the bar (bar 
strength and relative length) and the starburst activity in the center of 
barred spiral galaxies has been investigated by Chapelon, Contini \& Davoust 
(in preparation). 

Finally, our data have been used together with another sample of starburst 
galaxies to shed new light on the formation and chemical evolution of 
galaxies, by showing that SBNGs are still in the process 
of formation because of their underabundance in oxygen with respect to 
``normal'' spiral galaxies (\cite{COZIOLetal97a}1997a). 

\begin{acknowledgements}
Data from the literature were obtained with the Lyon Meudon Extragalactic 
database (LEDA), supplied by the LEDA team at CRAL-Observatoire de Lyon 
(France).  We thank Roger Coziol for helpful comments on the manuscript and 
the staff of Observatoire de Haute-Provence for assistance at the telescope. 
\end{acknowledgements}

\end{document}